%% file: main.tex
\documentclass[10pt]{article} 
\usepackage[preprint]{tmlr}

\input{math_commands.tex}

\usepackage{hyperref}
\usepackage{url}
\usepackage[pdftex]{graphicx}
\usepackage{booktabs}
\usepackage{datetime}

\newcommand{\subfigimg}[3][,]{%
  \setbox1=\hbox{\includegraphics[#1]{#3}}
  \leavevmode\rlap{\usebox1}
  \rlap{\hspace*{-6pt}\raisebox{\dimexpr\ht1-1\baselineskip}{#2}}
  \phantom{\usebox1}
}

\title{How Hard Is Quantum Advantage? A Cloud Microphysics Stress Test for Variational Quantum Models}

\author{\name Felix Herbort \email felix.herbort@planqc.eu \\
\addr PlanQC GmbH, Münchener Str. 34, 85748 Garching, Germany \\
\addr University of Hamburg, Institut für Quantenphysik, Luruper Chaussee 149, 22761 Hamburg,  Germany
\AND 
\name Ellen Sarauer \\
\addr Deutsches Zentrum für Luft- und Raumfahrt e.V. (DLR), Institut für Physik der Atmosphäre, Oberpfaffenhofen, 82234 We\ss ling, Germany
\AND
\name Daniel Ohl de Mello \\
\addr d-fine GmbH, An der Hauptwache 7, 60313 Frankfurt, Germany
\AND
\name Paul Christiansen \\
\addr d-fine GmbH, An der Hauptwache 7, 60313 Frankfurt, Germany
\AND
\name Steffen Hien \\
\addr d-fine GmbH, An der Hauptwache 7, 60313 Frankfurt, Germany
\AND
\name Cedric Brügmann \\
\addr d-fine GmbH, An der Hauptwache 7, 60313 Frankfurt, Germany
\AND
\name Dieter Jaksch \\
\addr University of Hamburg, Institut für Quantenphysik, Luruper Chaussee 149, 22761 Hamburg, Germany
\AND
\name Veronika Eyring \\
\addr Deutsches Zentrum für Luft- und Raumfahrt e.V. (DLR), Institut für Physik der Atmosphäre, Oberpfaffenhofen, 82234 We\ss ling, Germany \\
Institute of Environmental Physics (IUP), University of Bremen, Germany
\AND
\name Martin Kiffner \\
\addr PlanQC GmbH, Münchener Str. 34, 85748 Garching, Germany
\AND
\name Mierk Schwabe \\
\addr Deutsches Zentrum für Luft- und Raumfahrt e.V. (DLR), Institut für Physik der Atmosphäre, Oberpfaffenhofen, 82234 We\ss ling, Germany}

\def\month{07}  
\def\year{2026} 
\def\openreview{\url{https://openreview.net/forum?id=XXXX}} 

\begin{document}

\maketitle

\begin{abstract}
Quantum machine learning (QML) could have the potential to leverage advantages of quantum over classical computing but still lacks strong evidence of actual improvements and scalability, partly due to phenomena such as barren plateaus.
In this paper, we employ a hybrid quantum neural network (QNN) on a dataset on cloud microphysics, containing processes for phase transitions of water in the atmosphere and its related temperature changes, which are highly relevant for accurate climate predictions and projections.
To reach optimal performance of our QNNs, we employ a rich and trainable frequency spectrum together with expressivity enhancing classical postprocessing.
We find that our QNNs strongly benefit from extensive hyperparameter optimization and thereby demonstrate the feasibility of applying QNNs to complex physical systems.
At the same time, the QNNs are outperformed by classical baselines in the form of simple fully-connected neural networks.
We discuss identified bottlenecks of this class of quantum models to learn the full complexity of the cloud microphysics dataset to show that there is a need to further understand and improve variational quantum models for machine learning such that they might fill the gap where classical models fail or are inefficient.
\end{abstract}

\section{Introduction\label{sec:intro}}
Quantum computers are becoming more capable with increased gate fidelities and qubit numbers \citep{lib2026,gyger2024,bluvstein2024}. Also, the efficiency of quantum error correction is being improved so that algorithms that were thought to be decades away now seem within reach within reasonable timeframes \citep{googlequantumaiandcollaborators2025,webster2026,cain2026}.

Among many potential use cases for quantum computing devices, quantum machine learning holds a controversial stand.
Quantum neural networks with data reuploading have been shown to facilitate Fourier models with exponentially many modes \cite{shin_exponential_2023,schuld_effect_2021} and while exponentially large Hilbert spaces of quantum systems hold a strong potential for highly complex and parallelized data processing, this comes at the cost of flat gradient landscapes \citep{cerezo_does_2023}.
In addition, feeding unstructured, random classical data in and out of a quantum system can take exponential time in the number of qubits.
Nevertheless, for structured data, progress has been made for certain problems, for instance by employing matrix product state representations of data \citep{jaderberg2025,szoldra2026}.

At the same time, various fields are beginning to explore potential use cases of quantum machine learning, including weather and climate modelling \citep{tennie_quantum_2023,Jaderberg2024a,Schwabe2025a,Silva2025}.
For instance, clouds, and specifically cloud-aerosol interactions, remain one of the largest sources of uncertainty in climate projections  \citep{intergovernmental_panel_on_climate_change_ipcc_earths_2023}.
This is because cloud microphysics processes occur at subgrid scales, i.e., below the resolution of the climate model, and therefore need to be parameterized.
Machine learning has emerged as a tool to learn these parameterizations directly, either from same-resolution data for computational speed-up or from high-resolution data for accuracy improvements \citep{sarauer_physics-informed_2024,Lamb2026}.
\citet{Pastori2026} explored QML to predict cloud cover in the grid cells of a climate model, with a performance of the QML models comparable to that of classical NNs with the same number of free parameters.
However, predicting cloud cover is a computationally and physically simpler problem compared to cloud microphysics, where seven output features instead of one must be predicted and the underlying physics involves various phase transitions of water, driving changes in the respective phases' mass mixing ratios.

In this paper, we employ a combination of recent improvements to quantum neural networks (QNNs) in a single architecture on cloud microphysics, perform a hyperparameter optimization (HPO), and finally compare the results to a simple fully-connected neural network.
We show that QNNs, despite recent improvements and dedicated optimization, perform similarly to unoptimized neural networks on the dataset at hand.
However, optimized neural networks strongly outperform the QNN architecture.

We begin by describing our physical climate data generation process in Sec.~\ref{sec:data}.
Then, a detailed overview of data transformations, quantum and classical machine learning methods as well as HPO is presented in Sec.~\ref{sec:methods}.
We evaluate the HPOs in Sec.~\ref{sec:results} and discuss our results and their impact to the field in Sec.~\ref{sec:discussion} before closing with a brief summary and outlook in Sec.~\ref{sec:conclusion}.

\section{Data and preprocessing\label{sec:data}}
Our dataset involves cloud microphysics, specifically phase transitions of water that can occur due to environmental conditions such as air pressure or temperature.
However, simulating these processes by integrating the governing equations at spatial and temporal resolutions that are needed in climate simulations is computationally too expensive for current computing devices.
Therefore, physics-based parameterizations have been developed and optimized to approximate these processes \citep{stevens_what_2013}.
Still, simulating these processes relies on relatively high resolutions of about 5\,km per grid cell to resolve convective processes that are especially relevant for the projection of cloud processes while climate simulations have to rely on significantly coarser resolutions to deliver results in reasonable time.

We generate high-resolution simulation data using the non-hydrostatic atmospheric model ICON \citep{Zaengl_et_al_2015}, configured similarly to the QUBICC experiment \citep{Giorgetta_et_al_2022}.
The simulation is initialized from ERA5 boundary conditions \citep{Hersbach_et_al_2020} and uses prescribed boundary conditions for sea surface temperature, sea ice concentration, spectral solar irradiance, and well-mixed greenhouse gases.
Aerosol effects are neglected. The simulation employs a global ICON grid at approximately 5$\,$km horizontal resolution and 45 vertical levels.
The time step is 40 seconds.
Radiation is computed with the RRTMGP scheme \citep{Iacono_et_al_2008} every 12 minutes, while vertical diffusion and microphysics are evaluated every 40 seconds.
The ICON-JSBACH land surface model \citep{Schneck_et_al_2022} is activated in “lite” mode. Model output is written every 12 minutes for a total simulation time of 3 hours. The simulation is conducted for April 2004.
We store instantaneous 3D atmospheric state variables, e.g. microphysical tendencies, that we use to train our hybrid QML parameterization framework.
After a spin-up period of 3 timesteps, we use data from selected time intervals for training and testing of the QML model.
As in \citet{Grundner_et_al_2022}, we coarse-grain high-resolution outputs to a lower-resolution grid of $\sim 80\,$km by conservative averaging of variables across grid cells.

We describe the current state of a coarse cell as $x\in \mathbb{R}^{N_x}$ and the changes towards the next time step of our simulation as $y \in \mathbb{R}^{N_y}$ with $N_x = 10$ and $N_y = 7$.
Inputs $x$ include the values of the current timestep for mass mixing ratios of the various phases, other environmental information such as temperature and air pressure as well as information about the grid cell's latitude.
Outputs $y$ include changes of mass mixing ratios and temperature towards the next timestep.
A detailed list of all inputs and outputs can be found in Tab.~\ref{tab:features} in the appendix.
Finally, we split the data into train and test datasets by dividing the globe into longitudinal stripes such that cells corresponding to 8 equidistant ranges of longitude are assigned to the test dataset.
We choose the width of the ranges such that this method results in a 80-20 split of training and test data.
In total, the dataset comprises about 73 million training samples and about 18 million test samples.
The goal of our machine learning task is then to find a parametrization $\mathcal{F}_\theta: \mathbb{R}^{N_x} \rightarrow \mathbb{R}^{N_y}$ such that $\mathcal{F}_\theta (x) \approx y$ for all $(x,y) \in \mathcal{D}$ where $\mathcal{D}$ is the dataset.

\section{Methods}
\label{sec:methods}
To format climate data into a bounded domain that can be interpreted by quantum models using rotation gates, we use a data transformation method, introduced in Sec.~\ref{sec:data_transformation}.
Then, we present our model architecture in detail in Sec.~\ref{sec:hvqc} followed by a short sketch of their classical counterparts in Sec.~\ref{sec:fcnn}.
We close this section with a description of our hyperparameter optimization methodologies in Sec.~\ref{sec:hpo}.

\begin{figure}
    \centering
    \includegraphics[width=0.9\linewidth]{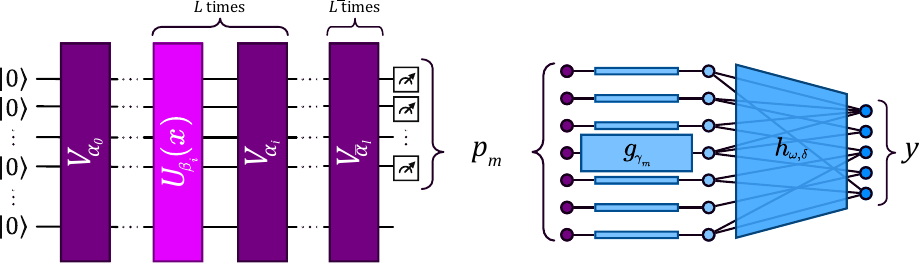}\llap{\parbox[b]{0.9\textwidth}{\textbf{(a)}\hspace{0.46\textwidth} \textbf{(b)\\\rule{0pt}{9em}}}}
    \caption{\label{fig:hvqc_model}Hybrid quantum neural network architecture. \textbf{(a)} Our model consists of a quantum circuit with repeated trainable data encoding layers $U$ and trainable entangling layers $V$. The circuit's output state is measured on $M$ qubits and basis state probabilities $p_m$. \textbf{(b)} The probabilities are then fed into a classical post processing pipeline where polynomials $g$ with trainable coefficients are applied to each $p_m$ and a weighted average $h$ computes the final output.}
\end{figure}

\subsection{Data transformations\label{sec:data_transformation}}
To normalize the inputs to the domain $[-\pi,\pi]$ in the dataset as well as potential new samples, which corresponds to the period of the $R_X$ gate, we scale the data using different data transforming techniques.
Our preprocessing method applies a tanh function to standardized inputs,
\begin{equation}
    \mathcal{T}^{\mathrm{A}}_{x_i} (vx_i) = \pi \tanh{\left(\mu_i\frac{x_i-\langle x_i\rangle}{\sqrt{\Delta x_i}}\right)}, 
\end{equation}
where $\langle \cdot \rangle$ denotes the mean, $\Delta x_i$ is the variance of $x_i$ and $\mu_i$ is a scaling hyperparameter.
For the outputs, we simply apply a standard scaler.

\subsection{Hybrid quantum neural network\label{sec:hvqc}}
Our model uses a hybrid structure where classical data is processed by a quantum system which is then measured as shown in Fig.~\ref{fig:hvqc_model} (a).
The measurement outcomes are fed into classical post-processing operations as depicted in Fig.~\ref{fig:hvqc_model} (b) to obtain the final result.

The quantum circuit consists of $L$ data encoding layers $U_{\beta_i}(x)$ where $\beta_i \in \mathbb{R}^{N_x}$ are trainable weights and $i \in \left\{0,\dots,L-1\right\}$ is the layer index.
Each $U_{\beta_i}(x)$ embeds data via angle encoding using single-qubit rotational gates around the $X$-axis of the Bloch sphere as
\begin{equation}
    R_X^{(j)}(\beta_{ij}\tilde{x}_{ij})
\end{equation}
acting on qubit $j$ with rolled input $\tilde{x}_{ij} = x_{(i+j) \mathrm{mod} N_x}$.
Encoding layers are alternated with variational layers that are implemented as entangling layers $V_{\alpha_i}$, where three trainable rotations 
\begin{equation}
    \prod_{\kappa \in \{X,Y,Z\}}R_\kappa^{(j)}(\alpha_{ij\kappa})
\end{equation}
are performed on every qubit $j$ with $\alpha\in{\mathbb{R}^{L\times N \times 3}}$, followed by a circular cascade of CNOT gates.
This corresponds to the implementation of \texttt{StronglyEntanglingLayers} in Pennylane.
We also consider variational layers with a single rotation around $Z$ for our hyperparameter optimization.
Finally, we read out quantum information by performing quantum state tomography on the first $M$ qubits to get $Z$-basis state probabilities $p_m$ with $m\in \left\{0,...,2^M-1\right\}$.

\citet{schuld_effect_2021} showed that this model results in a Fourier model that can incorporate a dense spectrum and thereby fulfills universal approximation theorems for $L\rightarrow \infty$.
According to \citet{shin_exponential_2023}, a dense spectrum with optimal scaling for the number of modes can be achieved by choosing the encoding parameters $\beta$ as
\begin{equation}
    \beta_{ij} = 3^i.
    \label{eq:exp_encoding}
\end{equation}
In this way, the model incorporates $\mathcal{O}(3^L)$ modes in the Fourier series, enabling a high expressivity for learning tasks.
Therefore, we initialize the parameters as given by Eq.~\ref{eq:exp_encoding} and scale them by $3^{-L}$ for normalization.
Although this allows for a large but finite amount of modes this initialization approach does not guarantee that the modes are actually the exact modes that are best-suited for the learning problem at hand.
Hence, we also allow the model to tune its Fourier frequencies by making $\beta$ trainable as suggested by \citet{jaderberg_let_2023} but treat this feature as a hyperparameter that can be enabled or disabled.

Note that because of our data reuploading strategy, our model has $\mathcal{O}(3^L)$ degrees of freedom but only $\mathcal{O}(LN)$ trainable parameters.
We believe that this is a key reason why QNNs can have an advantage over classical models in certain learning tasks.
Building on this scaling, we observed in preliminary tests that adding more variational layers between data reuploading and measurement can increase expressivity significantly.
This is because the increased number of variational parameters improves the controllability of the modes created by data reuploading.
We refer to these $\bar{L}$ additional layers as $V_{\bar{\alpha}_l}$ with $l\in \left\{0,...,\bar{L}\right\}$ with trainable variables $\bar{\alpha} \in \mathbb{R}^{\bar{L}\times N \times 3}$.

As described above, the quantum part gives us basis state probabilities $p_m$ that can be fed into classical post-processing which is depicted in Fig.~\ref{fig:hvqc_model} (b).
For this, we first apply a polynomial $g_{\gamma}$ of degree $P$ as
\begin{equation}
    g_{\gamma}(p) = \sum_{k=0}^P \gamma_{k} p^k
\end{equation}
to each of the $2^M$ basis state probabilities with $\gamma \in \mathbb{R}^{(P+1) \times 2^M}$ as trainable weights \citep{liao_expressibility-enhancing_2023}.
This again increases the amount of available modes.
We denote the output of the polynomial functions as $p'=(p'_1, \dots ,p'_{2^M})$.
Finally, we compute a weighted average as
\begin{equation}
    h_{\omega,\delta}(p') = \omega p' + \delta
\end{equation}
with trainable parameters $\omega \in \mathbb{R}^{2^M \times N_y}$ and $\delta \in \mathbb{R}^{N_y}$.
This again increases trainability and expressivity \citep{liao_expressibility-enhancing_2023}.
On another note, the classical post-processing allows the model to output values beyond the probability domain of $p_m \in [0,1]$.
In summary, trainable parameters are $\theta=(\alpha, \bar{\alpha}, \beta, \gamma,\delta,\omega)$.

Training is performed by gradient descent algorithm with Adam optimizer \citep{kingma_adam_2017}.
The loss function is given by the mean squared error (MSE)
\begin{equation}
    \mathcal{L}_\theta = \frac{1}{|\mathcal{D}|}\sum_{(x,y)\in\mathcal{D}} \left| \mathcal{F}_\theta(x) - y \right|^2.
\end{equation}
As we use state vector simulation and assume an infinite number of shots to produce our QNN results, we can obtain gradients via backpropagation.

\subsection{Classical neural networks}
\label{sec:fcnn}
To benchmark our QML ansatz, we also train classical fully-connected neural networks (FCNNs) with similar numbers of trainable parameters as in the QNN models.
FCNNs have already been shown to deliver accurate surrogate models for microphysics parameterizations for ICON \citep{sarauer_physics-informed_2024}.
Here, we consider FCNNs with $L_H$ hidden layers, containing $k_l$ kernels in layer $l$ with $l=0, ..., L_H-1$.
Each hidden layer applies an activation function $\phi$ but no activation is applied to the output.
Other modalities of the classical pipeline, including gradient calculation via backpropagation, preprocessing, and evaluation, are identical to our QNN.

\subsection{Hyperparameter optimization\label{sec:hpo}}
In sections \ref{sec:data_transformation}, \ref{sec:hvqc}, and \ref{sec:fcnn} we introduced multiple hyperparameters such as $L$, $\bar{L}$, whether to train $\beta$, and more.
These can influence model performance significantly and therefore require hyperparameter optimization (HPO).
One commonly used tool is Optuna \citep{akiba_optuna_2019}, which renders a generic library for arbitrary HPO tasks.
We perform a large amount of model trainings with hyperparameter values suggested by the tree-structured Parzen estimator (TPE) method after 15 randomly initialized warm-up trials \citep{bergstra_algorithms_2011,watanabe_tree-structured_2023}.
We provide a full list of the parameters in question in App.~\ref{app:hpo}.

Each model is trained for 200 epochs with 4096 training steps per epoch such that even with the minimum batch size $\min(N_B)=128$, the model trains on every sample in the dataset at least once.
Hence, increasing batch size only increases the amount of reuses of a sample for training.
Finally, we measure $R^2$ score performance on the test dataset in physical space, that is after applying the inverse preprocessing to the model's outputs, and use the average $R^2$ value over all target values as an objective for the HPO.

We perform one HPO for the QNN and FCNN architectures respectively in order to compare the presumably best model that can be obtained by employing each framework within the given frame of hyperparameters.

\section{Results\label{sec:results}}
In Sections \ref{sec:results-qnn} and \ref{sec:results-fcnn}, we will present the respective results of the performed HPOs of quantum and classical models. Then, we briefly compare the optimized models in Sec.~\ref{sec:results_q_vs_cl}.

\begin{figure}[t]
    \centering
    \includegraphics[width=\textwidth]{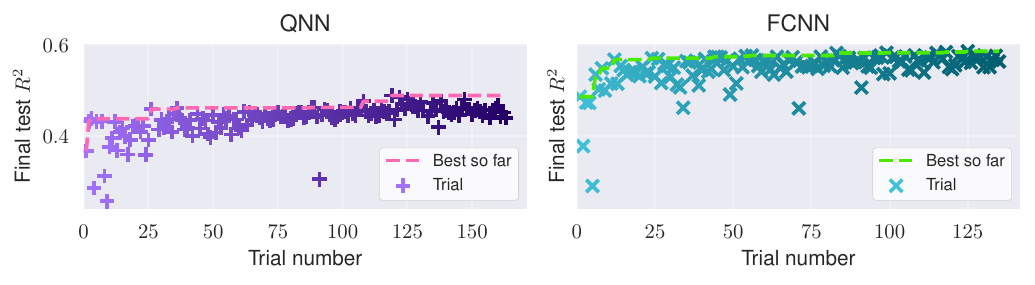}\llap{\parbox[b]{\textwidth}{\textbf{(a)}\hspace{0.48\textwidth} \textbf{(b)\\\rule{0pt}{10em}}}}
    \caption{Evolution of our hyperparameter optimization objective for our hybrid QNN in \textbf{(a)} and FCNN architecture in \textbf{(b)}. Each cross marks the $R^2$ test performance of a single HPO trial model. Color brightness additionally encodes trial number, consistently to figures below. A dashed line marks the best model's score up to the current state of the HPO.}
    \label{fig:hpo_progression}
\end{figure}

\subsection{QNN}
\label{sec:results-qnn}
We performed 157 model trainings in our HPO and present main results of the HPO here while additional information can be found in App.~\ref{app:hpo}.
Fig.~\ref{fig:hpo_progression} shows the evolution of the final test scores over the course of HPO.
After exhibiting a large spread of values at the beginning of the optimization, the averaged test $R^2$ converged to a value of approximately $0.48$ with the best QNN model exhibiting a score of $0.489$.
Except for few trials with significantly lower than optimal learning rate, most trials finished training with an $R^2$ score of at least $0.4$, highlighting the ability of QNNs to learn aspects of cloud microphysics with improved performance of optimized models.
We note that all models' scores presented in this paper exhibit sensitivity to far-outliers that we describe in App.~\ref{app:data_anomalies}.

\begin{figure}[t]
    \centering
    \includegraphics[width=\textwidth]{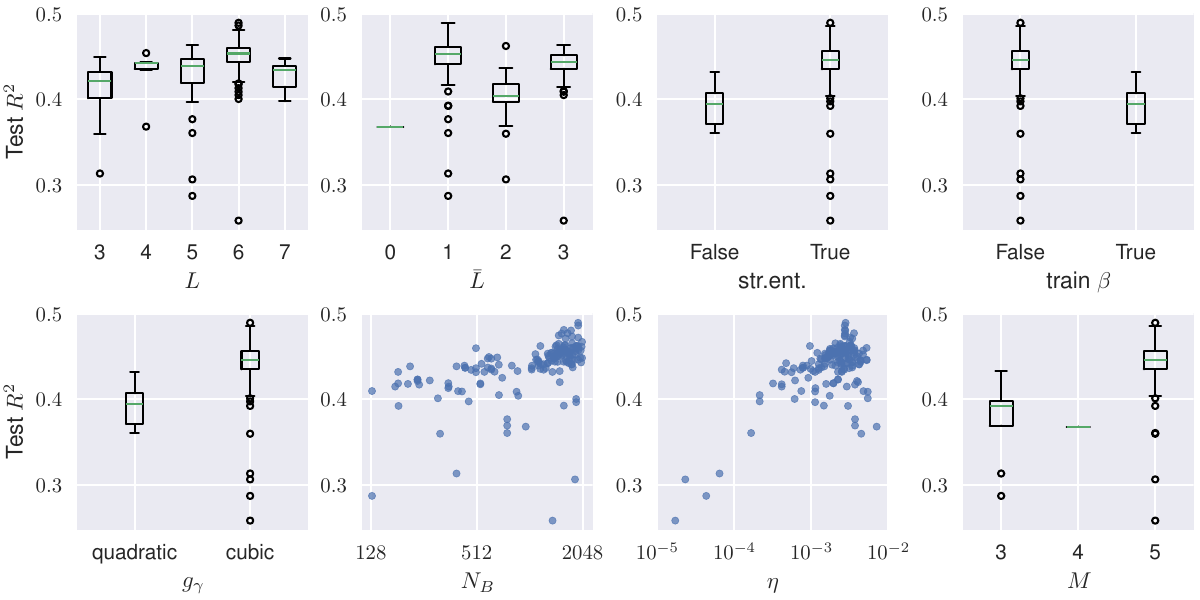}
    \caption{QNN performance slices for all hyperparameters throughout the HPO.
    The first row shows number of data reuploading layers $L$, additional variational layers $\bar{L}$, use of strongly entangling layers and whether to train reuploading factors $\beta$.
    The second row shows classical post-processing polynomial $g_\gamma$, batch size $N_B$, learning rate $\eta$, and number of measurement qubits $M$.
    Box-plots refer to discretized hyperparameters where in each box plot, the central green line denotes the median, the box spans the interquartile range (IQR, Q1–Q3), the whiskers extend to 1.5×IQR beyond the box edges, and individual dots mark outliers beyond the whiskers. For most hyperparameters, this data gives a clear guidance on how to choose parameters for QNN models for ICON microphysics data. However, strong improvement due to exponentially growing number of modes is not recognizable in the considered regime for the number of QNN layers $L$ and $\bar{L}$.}
    \label{fig:qnn_slice}
\end{figure}

According to the hyperparameter importances obtained via the fANOVA algorithm \citep{hutter_efficient_2014}, the most important hyperparameters for the QNN models were the number of measurement qubits $M$, batch size $N_B$, and learning rate $\eta$ (see Fig.~\ref{fig:app:importances} in the supplementary material).
Hyperparameters affecting the internal quantum architecture such as the number of layers $L$ and $\bar{L}$ however seem to be less relevant.
Moreover, Fig.~\ref{fig:qnn_slice} shows performances for slices through the hyperparameter space.
The plot shows slightly increasing performance with $L$ for $L\leq 6$ but $L=7$ already performs worse, which might be a first sign of circuit depth limiting trainability of QNNs due to barren plateaus.
Also appending at least one non-reuploading layer to the QNN increases performance significantly.
However, there is no straight-forward correlation beyond $\bar{L} \geq 1$.

As expected, introducing multiple trainable rotation gates by enabling strong entangling layers significantly improves the scores.
However, not training $\beta$~--~that is simply leaving the frequencies as initialized by the scheme by \citet{shin_exponential_2023}~--~results in increased performance.
This is an interesting result since giving the model fewer degrees of freedom and restricting the learning of frequencies of Fourier modes actually helps the model to learn the dataset.
A plausible explanation for this phenomenon is that the loss function exhibits more local minima for a trainable $\beta$ than if $\beta$ is fixed and therefore the training has a higher probability to converge in a local minimum.
Furthermore, a large batch size is favourable which is reasonable since the data has long tails in its distribution.
With increased batch size, training batches match the full distribution more accurately.
We also note that the performance peaks at a learning rate of approximately 2.8e-3.
Lastly, increasing the number of measurement qubits $M$ results in better performance.
Note that this also increases the amount of classical postprocessing as for each quantum output, we train an individual polynomial as well as a weighted average of $2^M$ inputs.

\subsection{Classical Neural Network}
\label{sec:results-fcnn}

\begin{figure}[t]
    \centering
    \includegraphics[width=\textwidth]{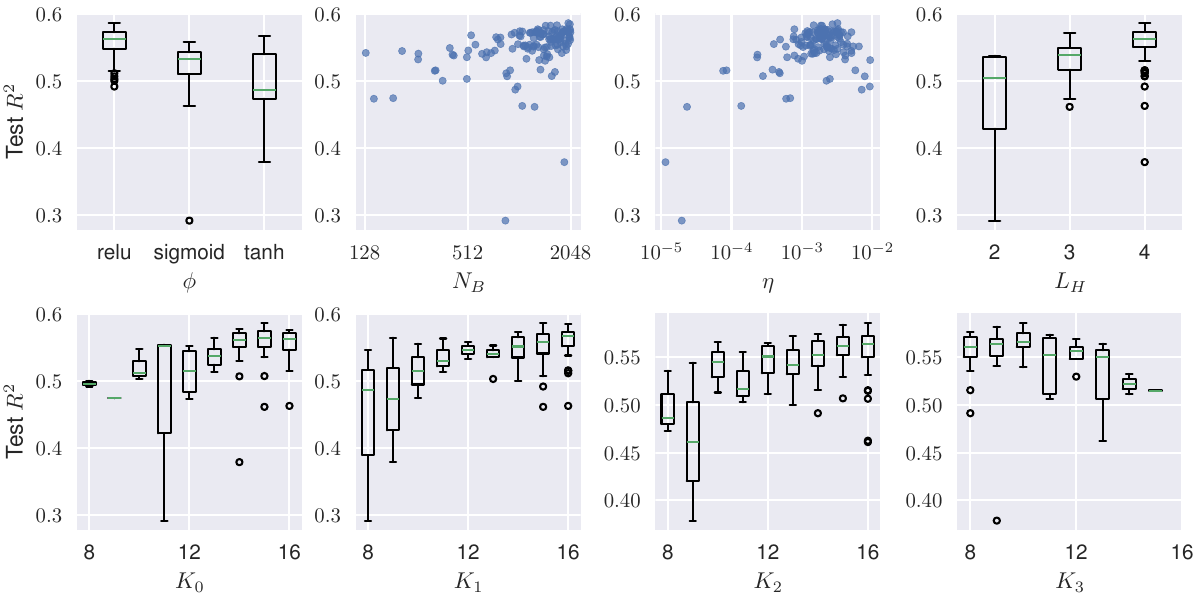}
    \caption{FCNN performance slices for all hyperparameters throughout the HPO.
    The first row shows activation function $\phi$, batch size $N_B$, learning rate $\eta$, and number of hidden layers $L_H$.
    The bottom row shows number of kernels $K_l$ for the respective hidden layer $l$.
    Box-plots refer to discretized hyperparameters where in each box plot, the central green line denotes the median, the box spans the interquartile range (IQR, Q1–Q3), the whiskers extend to 1.5×IQR beyond the box edges, and individual dots mark outliers beyond the whiskers. The optimized hyperparameters show good agreement with established standards in classical neural network literature. For instance, increasing the numbers of layers and kernels increases performance.}
    \label{fig:fcnn_slice}
\end{figure}

From Fig.~\ref{fig:hpo_progression} (b), we can deduce that the HPO for the FCNN architecture converged to $R^2$ value of approx. $0.56$ with best trial $0.585$.
Here, the most important hyperparameters are activation function $\phi$ and learning rate $\eta$. The batch size $N_B$ and the number of kernels in the first layer $K_0$ also have significant impact.
Referring to the slice plots in Fig.~\ref{fig:fcnn_slice}, correlations of hyperparameters and FCNN performance are rather intuitive.
For instance, increasing the number of kernels and layers improves performance. Only $K_3$ benefits from a more narrow architecture. This is a common choice for FCNNs where the network first creates a rich feature representation and then funnels information into a more dense representation for output inference.
Analogously to the quantum model, we find that larger batch sizes and a learning rate of around 2e-3 optimize the performance of the FCNN.

\subsection{Quantum vs classical}
\label{sec:results_q_vs_cl}
\begin{figure}[th]
    \centering
    \subfigimg[height=16em]{\textbf{(a)}}{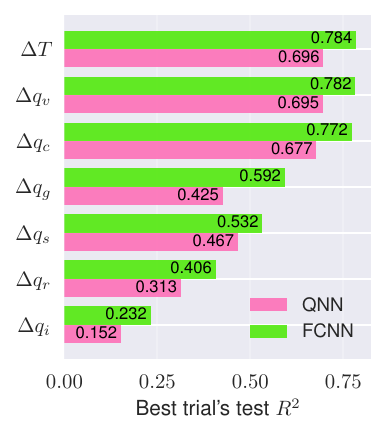}
    \hspace{1em}
    \subfigimg[height=16em]{\textbf{(b)}}{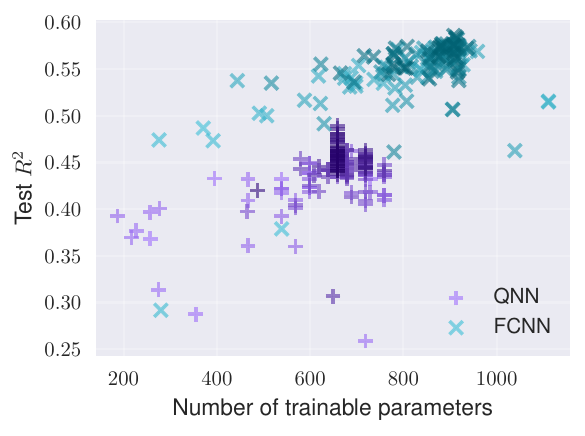}
    \caption{
        \textbf{(a)} $R^2$ scores of all targets for QNN and FCNN. For all models we have a score of $R^2 > 0$ but FCNN outperforms the best QNN for every output.
        \textbf{(b)} Effect of number of trainable parameters on model performance. Except for few outliers which arise from uncommon choices of hyperparameters, FCNNs consistently outperform the QNN model. However, we note that simple parameter count comparison should not be the only quantity to compare as we discuss in Sec.~\ref{sec:discussion}.
    }
    \label{fig:qnn_vs_fcnn}
\end{figure}
Comparing the best model from QNN and FCNN architectures respectively, we see that the FCNN outperforms the QNN not only in the averaged $R^2$ score but in each output individually as well (see Fig.~\ref{fig:qnn_vs_fcnn} \textbf{(a)}).
In Fig.~\ref{fig:qnn_vs_fcnn} \textbf{(b)}, we see that FCNNs systematically outperform QNN models, independently of parameter count (except for two instances which had a very low learning rate).
In fact, there are even FCNNs that outperform the best QNNs with $>30\%$ fewer trainable parameters.
Furthermore, note that FCNNs at the beginning of the HPO, i.e. with randomly chosen hyperparameters, are already performing just as well as the optimized quantum models.

\section{Discussion\label{sec:discussion}}
Our results clearly show the limitations of QNNs in the context of cloud microphysics:
We measure worse performance in terms of $R^2$ scores across the board.
Also the scalability of the quantum models is unclear since the performance drops after increasing model size beyond approximately 650 parameters.
However, we cannot rule out that the optimal architecture lies outside these ranges.
In fact, as long as the learning rate is of suitable order of magnitude, FCNNs do not strictly require hyperparameter optimization to beat QNNs.
On the other hand, our QNNs benefit from a hyperparameter optimization~--~a tool that is rarely used in QML literature.

Even more, although we are comparing similar numbers of trainable parameters, training a quantum neural network is significantly more expensive than training a neural network: The simulated training of the best QNN required close to 18 hours while the FCNN training took less than 5 minutes - both on NVIDIA H100 GPUs.
Performing such a training on a modern quantum device would take even longer because of slower quantum gate operations and repeated execution for expectation value estimation.

Comparing models of similar size in terms of the number of parameters does not necessarily give us information about models with significantly more parameters due to the lack of theoretical understandings of both QNN and FCNNs \citep{mingard_exploiting_2024}.
For FCNNs, we know because of numerous empirical studies that large models require additional effort like regularization techniques to train and the same will be necessary for QNNs.
However, we do not know yet which tools we need for QNNs to be efficiently trainable, although some tools are being developed \citep{kashif_deep_2025,hashizume_quantum_2026}.

Beyond the bare architectures compared here, classical ML microphysics schemes for this task rely on additional methods in preprocessing and training to achieve good performance, such as outlier enrichment to better cover extreme tendencies and physical constraints like a mass positivity condition added to the loss function \citep{sarauer_physics-informed_2025}.
These methods are not specific to classical networks and can equally be applied to quantum models.
This offers a clear perspective for improving QNN performance in future work. 

Furthermore, we note that our setup is rather experimental, looking only at offline models for cloud parameterizations.
For full operational use, one would need to couple such a model online into a climate model to see how it performs under realistic conditions.
Studies such as \citet{harder_physics-informed_2022} and \citet{Sarauer2026} have shown that strong offline performance does not always lead to strong online performance, which means that the quantum model would need to be tested online to see whether it is really worse when incorporated in an actual simulation.
A full online evaluation, however, is not within the scope of this study, since coupling a quantum model into a numerical climate model is a complex task left for future work. 

\section{Conclusion}
\label{sec:conclusion}
In this paper, we employed a hybrid quantum-classical model architecture to a cloud microphysics dataset, using a densely initialized but trainable spectrum and spectrum-enriching classical postprocessing.
We demonstrated that our quantum model shows basic expressivity and trainability by measuring the $R^2$ score.
However, even after performing a hyperparameter optimization, our quantum model does not outperform or even match classical fully-connected neural networks (FCNNs) with similar parameter counts.
Also, our analysis of parameter impact on model performance does not show strong signs of substantial improvement through increasing the number of layers as spectral analysis of QML models predicts \citep{shin_exponential_2023}.
On the other hand, FCNNs show potential for improvement in a rather straight-forward manner by increasing the network's width and depth.

This work shows that despite many claims of quantum advantage in recent literature \citep{abbas_power_2021,huang_quantum_2022}, variational quantum circuits can struggle when applied to real-world problems.
Nevertheless, climate modeling and other data-heavy computational tasks are in the need of more accurate and efficient parameterizations.
Strong advances in quantum machine learning research have to be made in order to overcome problems such as classical data input and output as well as vanishing gradients due to the curse of dimensionality.

\citet{zhao2026} recently introduced quantum oracle sketching and interferometric classical shadows, which offer provable memory advantages for loading streamed classical data and reading out low-dimensional functionals of the resulting quantum state.
A promising direction is quantum-accelerated reduced-order modeling: extracting the leading principal components of large climate datasets as classical shadows and estimating a compact linear propagator between them, yielding a low-dimensional surrogate in the spirit of dynamic mode decomposition.
Such a scheme would be inherently linear and restricted to a few spectrally well-separated modes, but it circumvents the classical input/output bottlenecks identified above and could compress climate datasets too large for classical memory into a compact, reusable surrogate.

\section*{Acknowledgments}

This project was made possible by the DLR Quantum Computing Initiative and the Federal Ministry of Research, Technology and Space; \href{https://qci.dlr.de/projects/klim-qml}{https://qci.dlr.de/projects/klim-qml}.

V.E. was supported by the Deutsche Forschungsgemeinschaft (DFG, German Research Foundation) through the Gottfried Wilhelm Leibniz Prize awarded to V.E. (Reference No. EY 22/2-1).
The authors also gratefully acknowledge the Earth System Modelling Project (ESM) for funding this work by providing computing time on the ESM partition of the supercomputer JUWELS \citep{JUWELS} at the Jülich Supercomputing Centre (JSC) for the simulation of the training data.
For the training of quantum and classical machine learning models, the HPC-cluster Hummel-2 at University of Hamburg was used. The cluster was funded by Deutsche Forschungsgemeinschaft (DFG, German Research Foundation) – 498394658.

DJ acknowledges support from the Hamburg Quantum Computing Initiative (HQIC) project EFRE. The EFRE project is co-financed by ERDF of the European Union and by “Fonds of the Hamburg Ministry of Science, Research, Equalities and Districts (BWFGB)”. DJ acknowledges support from the Cluster of Excellence ‘Advanced Imaging of Matter’ of the Deutsche Forschungsgemeinschaft (DFG)- EXC 2056- project ID 390715994, the European Union’s Horizon Programme (HORIZON CL42021- DIGITALEMERGING-02-10) Grant Agree ment101080085QCFD,DFGproject ‘Quantencomputing mit neutralen Atomen’ (JA1793/1-1, Japan-JST DFG-ASPIRE2024), and funding from the Federal Ministry of Research, Technology and Space (BMFTR)under the grant BeRyQC.

\bibliography{references}

\bibliographystyle{tmlr}

\include{appendix}
\end{document}

%% file: math_commands.tex
\usepackage{amsmath,amsfonts,bm}

\def\eqref#1{equation~\ref{#1}}

\def\1{\bm{1}}

\DeclareMathAlphabet{\mathsfit}{\encodingdefault}{\sfdefault}{m}{sl}
\SetMathAlphabet{\mathsfit}{bold}{\encodingdefault}{\sfdefault}{bx}{n}



%% file: appendix.tex
\appendix
\section{Supplementary Material}
This appendix provides additional information about our work.
First, we provide a more detailed overview of the data in Sec.~\ref{app:data} before outlining our implementation in code in Sec.~\ref{app:implementation}.
Then, we provide additional information on our hyperparameter optimization in Sec.~\ref{app:hpo}.
In Sec.~\ref{app:data_anomalies}, we note an anomaly in the training and test data.
Finally, we close with a brief description of additional work that we conducted that led to this paper.

\subsection{Cloud microphysics data}
\label{app:data}

Tab.~\ref{tab:features} contains a list of in- and outputs of our models.
In addition to mass mixing ratios, temperature and air pressure, we also provide the model with atmospheric wind $\omega_A$ and latitude $\phi$.
The latter is included to allow for incorporation of the Earth's latitude-depending surface rotation speed and the resulting Coriolis force.

\begin{table}[h]
\caption{Overview of all input and output features of the ML microphysics model (mmr: mass mixing ratio). Table adapted from \citet{sarauer_physics-informed_2025} with changed $\omega_A$ and added $\phi$.  \label{tab:features}}
\begin{tabular*}{\textwidth}{@{\extracolsep{\fill}}lccc@{}}
\hline
\textbf{Type} & \textbf{Short name} & \textbf{Description} & \textbf{Unit} \\
\hline
Input & pf\_mig ($p$) & air pressure & Pa \\
Input & ta\_mig ($T$) & air temperature & K \\
Input & qv\_mig ($q_v$) & water vapor mmr & kg\,kg$^{-1}$ \\
Input & qc\_mig ($q_c$) & cloud liquid water mmr & kg\,kg$^{-1}$ \\
Input & qi\_mig ($q_i$) & cloud ice mmr & kg\,kg$^{-1}$ \\
Input & qr\_mig ($q_r$) & rain mmr & kg\,kg$^{-1}$ \\
Input & qs\_mig ($q_s$) & snow mmr & kg\,kg$^{-1}$ \\
Input & qg\_mig ($q_g$) & graupel mmr & kg\,kg$^{-1}$ \\
Input & wa ($\omega_A$) & atmospheric wind & m\,s$^{-1}$ \\
Input & lat ($\phi$) & latitude & rad \\
\midrule
Output & tend\_ta\_mig ($\Delta T$) & tendency of air temperature & K\,s$^{-1}$ \\
Output & tend\_qv\_mig ($\Delta q_v$) & tendency of water vapor mmr & kg\,kg$^{-1}$\,s$^{-1}$ \\
Output & tend\_qc\_mig ($\Delta q_c$) & tendency of cloud liquid water mmr & kg\,kg$^{-1}$\,s$^{-1}$ \\
Output & tend\_qi\_mig ($\Delta q_i$) & tendency of cloud ice mmr & kg\,kg$^{-1}$\,s$^{-1}$ \\
Output & tend\_qr\_mig ($\Delta q_r$) & tendency of rain mmr & kg\,kg$^{-1}$\,s$^{-1}$ \\
Output & tend\_qs\_mig ($\Delta q_s$) & tendency of snow mmr & kg\,kg$^{-1}$\,s$^{-1}$ \\
Output & tend\_qg\_mig ($\Delta q_g$) & tendency of graupel mmr & kg\,kg$^{-1}$\,s$^{-1}$ \\
\hline
\end{tabular*}
\end{table}

\subsection{Implementation}
\label{app:implementation}

We implement both hybrid QNN and FCNN model architectures in a custom Python framework based on JAX \citep{bradbury_jax_2018}.
For the simulation of quantum circuits, our framework interfaces Pennylane \citep{bergholm_pennylane_2018}.
This combination allows us to compile the training pipeline as a whole and enables simulation on high-end GPUs like NVIDIA H100, enabling speedup factors on the order of thousands compared to single-CPU implementations with native Pennylane.
Gradients are evaluated through backpropagation.
Since the cloud microphysics dataset contains a large number of samples, this speedup is strictly required to make our study possible.

\subsection{Hyperparameter optimization}
\label{app:hpo}

\begin{figure}[ht]
    \centering
    \includegraphics[width=0.9\linewidth]{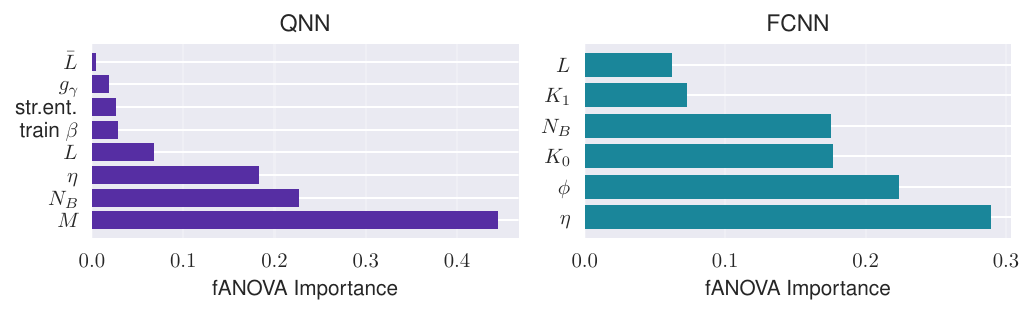}
    \caption{Hyperparameter importances for (a) QNN and (b) FCNN. For an explanation of the notations, see Tables~\ref{tab:hpo_qnn_parameters} and \ref{tab:hpo_fcnn_parameters} below.}
    \label{fig:app:importances}
\end{figure}

\begin{table}
    \caption{Allowed parameter ranges for HPO of QNNs.}
    \label{tab:hpo_qnn_parameters}
    \centering
    \begin{tabular}{|l|c|}
    \hline
    \textbf{Hyperparameter description} & \textbf{Range} \\ \hline
    learning rate $\eta$  & $[10^{-5},10^{-2}]$\\
    batch size $N_b$ & $[128,2048]$ \\
    number of reuploading layers $L$ & $[3,7]$\\
    additional variational layers $\bar{L}$ & $[0,3]$ \\
    number of measurement qubits $M$ & $[3,5]$ \\
    whether to train frequencies $\beta$& $\{\texttt{True},\texttt{False}\}$ \\
    number of rotations per layer & $\{1,3\}$ \\
    post-processing polynomial $g_\gamma$ & \{quadratic, cubic\} \\ \hline
    \end{tabular}
\end{table}

\begin{table}
    \caption{Allowed parameter ranges for HPO of FCNNs.}
    \label{tab:hpo_fcnn_parameters}
    \centering
    \begin{tabular}{|l|c|}
    \hline
    \textbf{Hyperparameter description} & \textbf{Range} \\ \hline
    learning rate $\eta$  & $[10^{-5},10^{-2}]$\\
    batch size $N_b$ & $[128,2048]$ \\
    number of hidden layers $L_H$ & $[2,4]$\\
    number of kernels $K_i$ in layer $i$ & $[8,16]$ \\
    activation function $\phi$ & \{relu, sigmoid, tanh\} \\ \hline
    \end{tabular}
\end{table}

Tables \ref{tab:hpo_qnn_parameters} and \ref{tab:hpo_fcnn_parameters} shows a list of the tuned hyperparameters and the ranges we considered for optimization of QNN and FCNN, respectively.
We decided for these ranges for reasons of sensible simulation times of QNN trainings and preceding experiments.
Restrictions for FCNN parameters were set to allow for similar numbers of parameters as the QNN model.
We intentionally allowed for slightly more parameters to enable a fair comparison, also regarding the more efficient execution of FCNNs on this scale.
Furthermore, Tables \ref{tab:top5_qnn} and \ref{tab:top5_fcnn} show the respectively 5 best trials from QNN and FCNN optimizations with the hyperparameters used.
Both HPOs converged to a stable model architecture. However, the FCNNs reach the limit of the allowed parameter range, indicating that the optimal model architecture might lie outside of the ranges shown in Tab.~\ref{tab:hpo_fcnn_parameters}.
Additionally, we provide the fANOVA hyperparameter importances in Fig.~\ref{fig:app:importances}.

\begin{table}
\caption{Top 5 QNN trials.}
\label{tab:top5_qnn}
\centering
\begin{tabular}{|r|rrr|rrrrrrrr|}
\hline
trial & train $R^2$ & test $R^2$ & \# weights & $L$ & $\bar{L}$ & $M$ & $N_\mathrm{B}$ & $\eta$ & $g_\gamma$ & train $\beta$ & str. ent. \\
\hline
118 & 0.466 & 0.489 & 659 & 6 & 1 & 5 & 1931 & 2.81e-3 & cubic & False & True \\
121 & 0.455 & 0.486 & 659 & 6 & 1 & 5 & 1933 & 2.74e-3 & cubic & False & True \\
124 & 0.457 & 0.481 & 659 & 6 & 1 & 5 & 1601 & 2.74e-3 & cubic & False & True \\
146 & 0.455 & 0.480 & 659 & 6 & 1 & 5 & 1760 & 2.79e-3 & cubic & False & True \\
137 & 0.446 & 0.477 & 659 & 6 & 1 & 5 & 1683 & 3.62e-3 & cubic & False & True \\
\hline
\end{tabular}
\end{table}

\begin{table}
\caption{Top 5 FCNN trials.}
\label{tab:top5_fcnn}
\centering
\begin{tabular}{|r|rrr|rrrrrrrr|}
\hline
trial & train $R^2$ & test $R^2$ & \# weights & $L_H$ & $K_0$ & $K_1$ & $K_2$ & $K_3$ & $N_B$ & $\eta$ & $\phi$ \\
\hline
124 & 0.555 & 0.586 & 908 & 4 & 15 & 15 & 16 & 10 & 1988 & 3.22e-3 & relu \\
117 & 0.553 & 0.584 & 913 & 4 & 15 & 16 & 15 & 10 & 2030 & 2.73e-3 & relu \\
98 & 0.560 & 0.583 & 913 & 4 & 15 & 16 & 15 & 10 & 1706 & 1.48e-3 & relu \\
114 & 0.556 & 0.582 & 913 & 4 & 15 & 16 & 15 & 10 & 1565 & 1.86e-3 & relu \\
82 & 0.557 & 0.582 & 916 & 4 & 15 & 16 & 16 & 9 & 1465 & 1.94e-3 & relu \\
\hline
\end{tabular}
\end{table}

\subsection{Data anomalies}
\label{app:data_anomalies}
Throughout the training of our models, we observe a systematically better score on test data than on training data.
Generally, this can only appear if either the implementation is faulty or if there are structural differences between training and test data.
After careful investigation of our code and the data, we track the reason for the discrepancy back to few extreme outliers in the training data that could not be estimated well by any of our models.
These extreme outliers are more than $100$ standard deviations away from the mean value, sometimes resulting in an MSE of $O(10^4)$, which can have significant impact on the mean over all samples in the dataset.

To overcome this problem, we would need either even more data than we currently have, split the data evenly such that the distributions are more similar, or remove the outliers, which would not be desirable since we need to learn extreme weather events for accurate climate simulation.
Also, we can not decrease the number of training samples as this would decrease overall model performance.
Hence, we simply accept the lower training score and note that this is due to both models not being able to generalize to extreme weather events.

\subsection{Additional work}
\label{app:further_work}
Throughout the project, we employed a larger variety of methods than shown in this paper.
First, we analyzed the effect of various data transformations including standard scaling, Gaussian and uniform probability integral transforms, logarithmic transformations designed for atmospheric processes \citep{Pastori2026} as well as combinations thereof.
Furthermore, we employed functions transforming the argument of input encoding rotational gates as proposed in \citet{liao_expressibility-enhancing_2023}. However, we believe based on preliminary results that the exponential mode scaling as described by \citet{shin_exponential_2023} has a larger effect on model expressivity than introducing additional basis functions.
As an alternative architecture, we also employed hybrid quantum generative adversarial networks (QGANs) as a probabilistic alternative to the deterministic QNNs.
However, due to significantly longer training times, a full training of such a model was not feasible on cloud microphysics data.
Among these methods, the methods shown in this paper were showing best performance in early tests.
Hence, we restricted the HPO to the selected ones in order to perform the simulations in reasonable time.